\documentclass[pra,twocolumn,amsmath,amssymb,floatfix]{revtex4}
\usepackage{graphicx}% Include figure files
\usepackage{ulem}
\usepackage{color}

\newcommand{\be}{\begin{eqnarray}}
\newcommand{\ee}{\end{eqnarray}}

\newcommand{\wbe}{\begin{widetext}}
\newcommand{\wee}{\end{widetext}}

\begin{document}

\title{Quantitative Studies on the Critical Regime near Superfluid to Mott Insulator Transition}

\author{Hao Lee$^{1,2}$, Shiang Fang$^{3}$, Daw-Wei Wang$^{1,2}$}

%\email{s100022806@m100.nthu.edu.tw}

\affiliation{$^{1}$ Physics Department, National Tsing Hua University, Hsinchu,
Taiwan
\\
$^{2}$ Physics Division, National Center for Theoretical Sciences,
Hsinchu, Taiwan
\\
$^{3}$ Department of Physics, Harvard University, Cambridge, Massachusetts 02138, USA
}

\date{\today}

\begin{abstract}
We investigate the critical behaviors of correlation length and critical exponents for strongly interacting bosons in a two-dimensional optical lattice via quantum Monte Carlo simulations. By comparing the full numerical results to those given by the effective theory, we quantitatively determine the critical regime where the universal scaling behaviors applies at a finite temperature, for both classical Berezinskii-Kosterlitz-Thouless transition and quantum phase transition from superfluid to Mott insulator. Our results represent the critical regime that should be observed in present experimental conditions.
\end{abstract}

\maketitle
%%%%%%%%%%%%%%%%%%%%%%%%%%%%%%%%%%

%%%%%%%%%%%%%%%%%%%%%%%%%%%%%%%%%%%\section{Introduction}

%GENERAL INTRO.% 
%\underline{\textit{Introduction}} 
\section{Introduction}

The exotic many-body phases and their phase transition properties have been extensively investigated in the systems of ultracold atoms in recent decades, mostly due to the defect-free and widely tunable characters in the real experiments \cite{ReviewBloch1, ReviewBloch2, ReviewChin, MetIns, BlochNature2000, JochimPRL2015, CMC}.
Besides of quantum phase diagrams and transition boundaries, more and more attention are drawn to investigate the critical properties near the phase transition points \cite{Laloe, Smith, Ohashi, Mueller, ZhouHo, VicariPRB14, Holzmann, NavonScience, ProkofevPRL14, ChinNature11, Esslinger_Science, ChinCriticalityNJP}, because finite temperature and finite size effects are very important and unavoidable in a cold atom experiment. From theoretical points of view, critical regime is mainly controlled and determined by the effective theory near the transition point, while the validity of the effective theory has to be determined only by higher order effects or full numerical simulations.

Taking interacting Bose gases as examples, thermal fluctuations in low dimensional system may lead to interesting topological excitations and the Berezinskii-Kosterlitz-Thouless transition in two-dimensional systems \cite{BKT, PathIntegral1987, Gora_QC, Cooper, Laloe, JSY_aniso}, which has been realized in current experiments \cite{Hadzibabic, Phillip_QCvsSFexp, Ian_PRL105}. When loading into an optical lattice, interacting bosons may undergo a superfluid (SF) to Mott insulator (MI) transitions \cite{Spielman_CondensateFraction, Chin_Nat} as predicted from a single band Bose-Hubbard model \cite{FisherFisher}. Some investigations on the critical behaviors have been carried out both experimentally \cite{Esslinger_Science, NavonScience, Polkovnikov_PNAS, Endres} and theoretically \cite{ChinCriticalityNJP, Lode_QMC, Fang_QMC, Rancon}. 
However, how to quantitatively determine the critical regime and how the classical phase transition at finite temperature has a crossover to quantum critical regime near zero temperature is still unclear.

In this paper, we quantitatively identify the Berezinskii-Kosterlitz-Thouless (BKT) regime and the quantum critical (QC) regime in a two-dimensional Bose-Hubbard model by comparing numerical results from the {\it ab initio} path-integral quantum Monte Carlo (QMC) \cite{PathIntegral1987, Prokofev1, Prokofev2} with results from the effective theories. More precisely, we define a proper approach to investigate how the correlation length ($\xi$) and the critical exponent ($\eta$) changes as a function of system parameters near the transition boundaries. 
The finite temperature crossover regime has an asymmetric shape in the superfluid side and in the disordered side, showing a non-trivial higher order correction due to different ground states. Finally we discuss the correlation length as well as the critical exponents can be determined in a finite-size system. 

In Sec. II, we first introduce the model to study and the critical behaviour of correlation function near the phase transition point. In Sec. III, we systematically investigate the signature of the correlation function near the BKT transition point, and extend it to the quantum critical regime of the SF-MI multicritical point in Sec. IV. Finally, we discuss the experimental application in Sec. V.
 
%%%%%%%%%%%%%%%%%%%%%%%%%%%%%
\section{System Hamiltonian and correlation function}
%%%%%%%%%%
\subsection{General description of the phase diagram}

%------------
In order to quantitatively investigate how the classical BKT phase transition between 2D superfluid and normal sate merges with Mott insulator (MI) phase at a quantum critical point, we consider bosonic ultracold atoms loaded in two-dimensional optical lattice, which is known well-described by the single-band Bose-Hubbard model:
%-----------
\begin{equation}
H=-J\sum_{\left\langle i,j \right\rangle} (a^{\dagger}_{i}a_{j}+h.c.)+\frac{U}{2}\sum_{i}n_{i}(n_{i}-1)-\mu\sum_{i}n_{i},
\end{equation}
%-------------
where $a_{i}$ and $n_{i}$ are bosonic field operator and density operator respectively at the $i$th lattice site, $J$ is the nearest-neighbor hopping energy, $U$ is the on-site interaction, and $\mu$ is the chemical potential. The phase diagrams at zero temperature and finite temperature have been investigated before \cite{FisherFisher, Fang_QMC}. 
%In Fig. \ref{fig:1}, we show a schematic plot of the zero temperature and finite temperature phase diagram together. 
We also show quantum critical regime as well as the critical regime of BKT transition. We will then give a more concrete study in the rest of this paper. 

%%%%%%%%%%
\subsection{The critical properties of correlation function}

In the previous works of quantum Monte Carlo calculation, the phase boundary between superfluid and normal state (as well as MI state at zero temperature) is determined by the superfluid density (or winding number) \cite{PathIntegral1987}.
However, this method cannot be used to determine the quantum critical regime, which depends on the effective model in a certain parameter regime. In our calculation below, we investigate quantum critical regime through the long-range behaviour of the correlation function, which is defined as following:
%--------------
\begin{equation}
G_{i,j}\equiv\left\langle a^{\dagger}_{i}a_{j}\right\rangle \propto r^{-(d-2+\eta)}\exp(-r/\xi) \ \ \ {\rm for}\ \ |i-j|\to\infty,
\label{eq:aa}
\end{equation}
%-----------
where $r=r_{j}-r_{i}$, $d=2$ is the dimension of the system, $\eta$ is a critical exponent and $\xi$ is the correlation length. It is known that in 3D systems, if $G_{i,j}$ goes to a non-zero constant value as $|i-j|\to\infty$, it indicates the superfluidity with an off-diagonal long-ranged order. However, the situation becomes much more tricky in lower dimensional system, which does not have a true long-ranged order at any finite temperature according to Mermin-Wagner theorem \cite{MerminWagner}.

In a 2D system with a weak interaction ($U\ll J$), we know that the system can undergoes a BKT transition to be a superfluid at a finite temperature $T_c$. The correlation function has an exponential decay, i.e. $G_{i,j}\sim\exp(-r/\xi)$ with a finite value of $\xi$, for $T>T_c$. The the critical exponent $\eta\sim 0.01$ is known very small and can be neglected \cite{Esslinger_Science}. However, when $T<T_{c}$, the binding of a vortex and an anti-vortex strongly suppresses the thermal fluctuations and therefore the correlation function shows a power-law decay with a divergent correlation length, i.e. $\xi\to\infty$ and $\eta$ has a finite value, i.e., $G_{i,j}\sim r^{-\eta}$. 

\begin{figure}[htbp!]
\centering
\includegraphics[width=8.5cm]{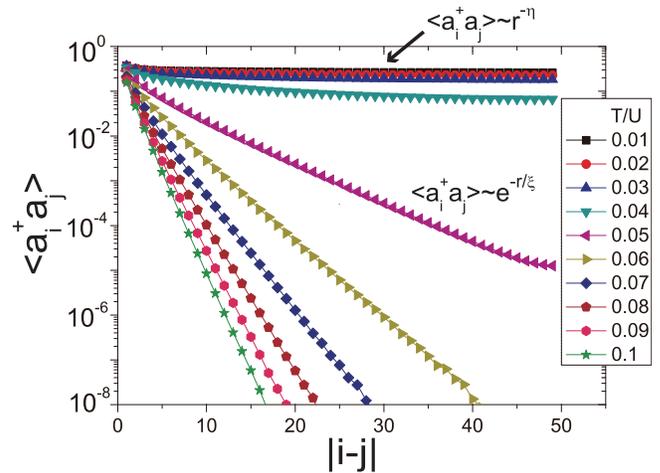}
\caption{The semi-log plot of the correlation function $G_{i,j}=\left\langle a^{\dagger}_ia_j\right\rangle$ as function of distance $r\equiv|i-j|$ where $i=0$ at $T/U=0.01$ to $0.1$ from up to low are shown at $J/U=0.04$ and $\mu/U=0$.
The periodic boundary condition with $100\times 100$ square lattice is applied. 
The exponential decay at $T/U=0.05$ to $0.1$ is shown here, while the algebraic decay at $T/U=0.01$ to $0.04$ can be identified in a log-log plot.
}
\label{fig:aa-r}
\end{figure}

In Fig. \ref{fig:aa-r}, we show how the correlation function $G_{i,j}$ changes for different temperatures with $J/U=0.04$ and $\mu/U=0$. Here for simplicity, we just show the results of a $100\times 100$ square lattice. Results of larger lattice sites are also obtained but did not show significant differences. As one can see that the correlation function changes significantly from an exponential decay to a power-law decay when temperature decreases. 
The critical temperature observed (between 0.04 and 0.03 $U$) from the change of correlation function is consistent with the value estimated ($T_c\sim 0.036 U$) from the winding number fluctuation, i.e. $\left\langle W^{2} \right\rangle=4/\pi$ \cite{JSY_aniso}. 
As we will show below, the quantum critical regime can be defined by how the correlation length, $\xi$, changes as a function of temperature and/or other system parameters.
%--------Added in resubmission-------
We also emphasize that in our numerical simulation, the correlation function $G_{i,j}$ is found not affected by the finite size effect except in the range $|i-j|\approx L/2$, e.g., as shown in the purple triangles in Fig. \ref{fig:aa-r}. Besides, the correlation length $\xi$ extracted from the $G_{i,j}$ in the regime above critical temperature is also well below the system size. Therefore, all the results presented in this paper should be well justified and applicable in the thermodynamic limit.
%-------------------------------------
%=================
%===================

%%%%%%%%%%%%%%%%%%%%%%%%
\section{Critical regime of Berezinskii-Kosterlitz-Thouless (BKT) regime}

Before investigating the quantum critical regime, it is necessary to study how the correlation length ($\xi$) or the critical exponent ($\eta$) changes when the temperature is approaching $T_c$ from the higher or lower temperature side. According to the BKT theory \cite{BKT, Zinn-Justin}, the unbound topological defects (i.e. vortices and anti-vortices) lead the correlation length $\xi$ to diverge as $\sim \exp\left[c\left(\frac{T-T_c}{T_c}\right)^{-1/2}\right]$ near the critical temperature, where $c$ is a dimensionless constant. 
%===============
\begin{figure}[htbp!]
\centering
\includegraphics[width=8.2cm]{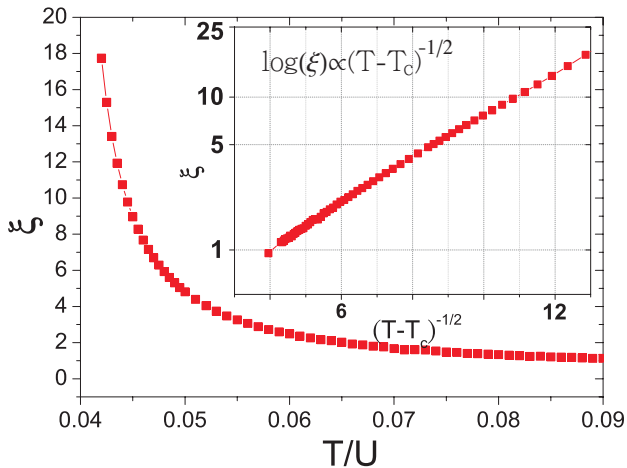}
\caption{
The correlation length $\xi$ as function of temperature, where $\xi$ is fitted as $G_{i,j}\propto \exp(-r/\xi)$ at $T$ above $T_c$, $J/U=0.04$ and $\mu/U=0$ in $100\times 100$ square lattice with periodic boundary condition.
In the inset, the semi-log plot of $\xi$ as function of $(T-T_{c})^{-1/2}$, the divergent law of $\xi$ as $T$ decreasing is shown to be $\log (\xi)\propto (T-T_{c})^{-1/2}$, which agrees with the one in BKT transition regime \cite{Zinn-Justin}. 
The $T_{c}$ is defined as the temperature that $\left\langle W^{2} \right\rangle=4/\pi$.}
\label{fig:xi}
\end{figure}
%===============

In Fig. \ref{fig:xi}, we show how the correlation function diverges as a function of temperature when $T>T_c$ for $J/U=0.04$ and $\mu/U=0$. As we can see from the inset that $\log\xi\propto (T-T_c)^{-1/2}$ as $T\to T_c$. However, when the temperature is higher then $T_c$ up to a certain value, we do find some deviation from the linear relation, indicating a higher order correction of the BKT theory. As a result, we can define another temperature scale, $T^{BKT}_\xi$, above which the BKT theory becomes not reliable upto certain accuracy. In this paper, we define $T^{BKT}_\xi$ such that the linear relationship between $\log\xi$ and $(T-T_c)^{-1/2}$ is accurate within 99\% for $T_c<T<T^{BKT}_\xi$.
We note that $T^{BKT}_\xi$ defined here is {\it not} another critical temperature, but just a temperature scale one can tell how much accurately the BKT effective theory can be quantitatively applied near the critical temperature. This temperature scale is very important to our future analysis.

When $T$ is below $T_{c}$, on the contrary, the correlation function decays algebraically, i.e., $G_{i,j}\propto r^{-\eta}$ as $r\to\infty$. Now we will compare two values of $\eta$. $\eta_{\rm num}$ is obtained from the numerical (QMC) calculation through fitting the slope of $G_{i,j}$ in the log-log plot (not shown here). $\eta_{\rm BKT}$ is derived from the BKT theory from the superfluid density, $\rho_s$, through $\eta_{\rm BKT}=mT/2\pi\rho_{s}\hbar^{2}$ \cite{Polkovnikov_PNAS}. In other words, $\eta$ becomes a universal value, $\eta_{c}=1/4$, as $\rho_{s}(T_{c})=2mT_c/\pi\hbar^{2}$ at phase transition point.

In our QMC calculation, the superfluid density can be calculated from the winding number, i.e., $\rho_{s}=\frac{m}{\hbar^{2}}\frac{\left\langle W^{2} \right\rangle L^{2-d}}{d\beta}$, where $m=1/(2J)$ in Bose-Hubbard model, $L$ is system size, $d$ is the dimensions, $\beta=1/k_{B}T$, and $\left\langle W^{2} \right\rangle$ is the mean of winding number square \cite{PathIntegral1987}. 
Ideally the critical temperature $T_c$ in thermodynamic limit ($L\rightarrow\infty$) is determined by the discontinuity of superfluid density and mean-squared winding number. In a finite 2D system, however, the mean-squared winding number $\left\langle W^2 \right\rangle$ still converges to $4/\pi$ at $T_c$ as the size being sufficiently large ($L\geq 100$) \cite{JSY_aniso, PathIntegral1987}. We have checked separately and confirm this results for various system sizes (not shown here). Therefore, $\eta_{\rm BKT}$ is also calculated from the BKT theory via the calculated superfluid density. Comparing the results of $\eta_{\rm num}$ and $\eta_{\rm BKT}$ can provide information about how accurate the BKT theory can be applied when near the critical temperature from below.

\begin{figure}[htbp!]
\centering
\includegraphics[width=8.5cm]{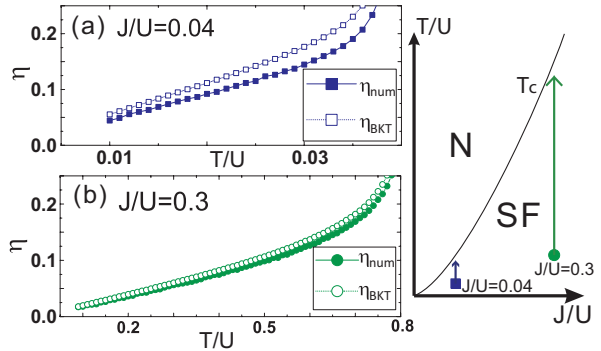}
\caption{The critical exponent $\eta$ as function of T plot at $\mu/U=0$, $J/U=0.04$ (a) and $0.3$ (b) respectively.
The filled marks with solid curves are the slopes from fitting from the log-log plot of the correlation function as $G_{i,j}\propto r^{-\eta_{\rm num}}$. The blanked marks with dot curves are $\eta_{\rm BKT}$ calculated from superfluid density. }
\label{fig:eta}
\end{figure}

In Fig. \ref{fig:eta}(a) and (b), the values of $\eta_{\rm num}$ (filled squares and circles) and $\eta_{\rm BKT}$ (blanked square and circles) are shown together as function of $T/U$ at $\mu/U=0$ for two different values of tunnelling amplitude, $J/U=0.04$ and $0.3$, respectively. On can see that both $\eta_{\rm BKT}$ and $\eta_{\rm num}$ increase up to $1/4$ as the system approaching to phase transition point, while $\eta_{\rm num}$ and $\eta_{\rm BKT}$ match each other quantitatively only when deep inside the superfluid regime (i.e. when temperature is in the lower side and/or $J/U$ is in the larger side.) Therefore, as expected, the deviation can be more significant when the system parameters are tuned close to the quantum critical point, where the critical temperature approaches to zero.
%===================
%=================

%%%%%%%%%%%%%%%%%%%%%%%%\section{D: nu & QCP}
\section{Crossover regime near the quantum critical point}

\begin{figure}[!hbp]
\centering
\includegraphics[width=8.5cm]{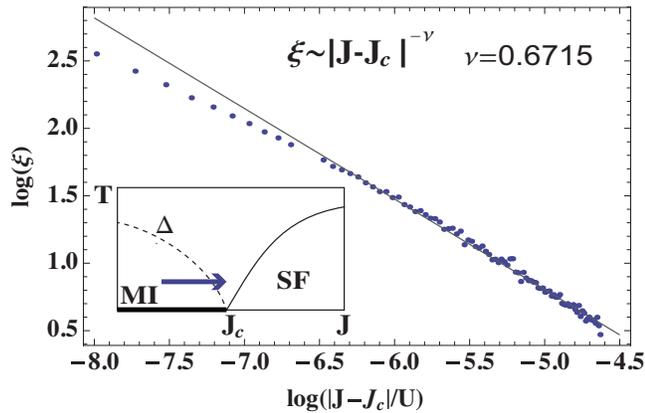}
\caption{The log-log plot of $\xi$ to $|J-J_c|/U$ %from $J/U=0.05$ to $0.0597$ 
at $\mu/U=0.37$ and $T/U=0.001$ in a $128\times128$ square lattices. The universality class of 3D-XY model is characterized by the relation $\xi\propto|J-J_c|^{-\nu}$, where $\nu\approx 0.6715$ is the correlation length exponent as the solid line.}

\label{fig:nufit}
\end{figure}

As approaching to the quantum critical regime, it is known that the Bose-Hubbard model in 2D square lattices shares the same universality class with the (2+1)D $XY$ model (i.e. the $O$(2) rotor model) in the vicinity of the multicritical point (the Mott lobe tip), because of the infinite integration range of the imaginary time axis \cite{Sachdev}. As a result, in the disordered side and along the zero temperature line (i.e. in the Mott insulator regime), the correlation length should diverge as $\xi\propto|J-J_c|^{-\nu}$ for $J\rightarrow J_c^-$,  where $\nu\approx0.6715$ is the correlation length exponent for $O$(2) rotor model and $J_c/U\approx0.05974$ for the first Mott lobe tip \cite{Barbara_nu}. 
In our numerical result as shown in Fig.(\ref{fig:nufit}), the correlation length at extremely low temperature ($T/U=0.001$) follows the such power-law behavior very well in the range $0.05<J/U<0.0596$. Finite temperature effects intervene when approaching the multicritical point further because the energy gap becomes too small. 
%--------------
%------------

\begin{figure}[!hbp]
\centering
\includegraphics[width=8.5cm]{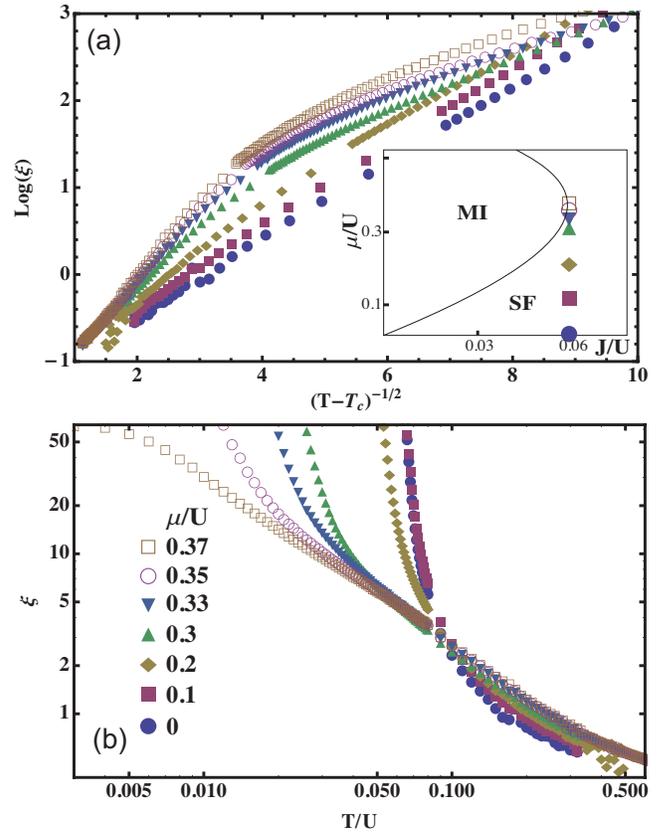}
\caption{(a) The semi-log plot of $\xi$ to $(T-T_{c})^{-1/2}$ at $J/U=0.05974$ in $128\times128$ square lattices. The inset shows the locations of the parameters in zero-temperature phase diagram. The BKT critical behaviors of correlation length, $\log (\xi)\propto (T-T_{c})^{-1/2}$ is linear in this plot.
The $\mu/U=0, 0.1, 0.2, 0.3, 0.33, 0.35, 0.37$, from bottom to top respectively. 
(b) The log-log plot of $\xi$ to $T$ at the same parameters as (a).
The quantum critical behaviors of correlation length, $\xi\propto 1/T$, is linear in this plot and well-followed as $\mu/U\to 0.37\sim \mu_c/U$.
}

\label{fig:crossover}
\end{figure}

In the other side of the critical point, the superfluid-normal phase transition is still of BKT type. In Fig. \ref{fig:crossover}(a), we show results for different values of chemical potential $\mu/U$ as $J=J_c$ for comparison. It is easy to see that the range of the BKT critical behavior, i.e., $\log (\xi)\propto (T-T_c)^{-1/2}$, shrinks when close to the quantum critical point ($T_c\to 0$) as expected. In our numerical results, this regime does not shrink to absolute zero because we have set 99\% (instead of 100\%) accuracy in the fitting criteria.
 
%--------------
%------------

Now we consider the finite temperature effect near the multicritical point, i.e., $(J,\mu)\rightarrow(J_c,\mu_c)$, the divergence of correlation length $\xi$ with respect to temperature is predicted to be universal, $\xi\propto 1/T$, in a large temperature interval, and the lower bound of this interval approaches $T=0$ at the multicritical point \cite{SachdevPRB}. 
Here we numerically explore the crossover regime by investigating how the correlation length changes with respect to the temperature in different parameter regime. As clearly shown in Fig. \ref{fig:crossover}(b), the quantum critical behavior, $\xi\propto 1/T$, is revealed and strongly deviates from BKT behavior as close to the quantum critical point (QCP).
Similar to previous definition of $T^{BKT}_{\xi}$ in Sec. III, we can then define another temperature scale, $T^{QC}_{\xi}$, such that in the temperature interval, $T^{QC}_{\xi}<T$, the correlation length fits the QC critical behaviour upto 99\% in accuracy. However, we have to mention that in an even higher temperature regime ($T\gg T_{\xi}^{QC}$, not shown in the figure), the classical fluctuation is expected to dominate, destroying the $\xi\propto 1/T$ behavior. 

\begin{figure}[!hbp]
\centering
\includegraphics[width=8.5cm]{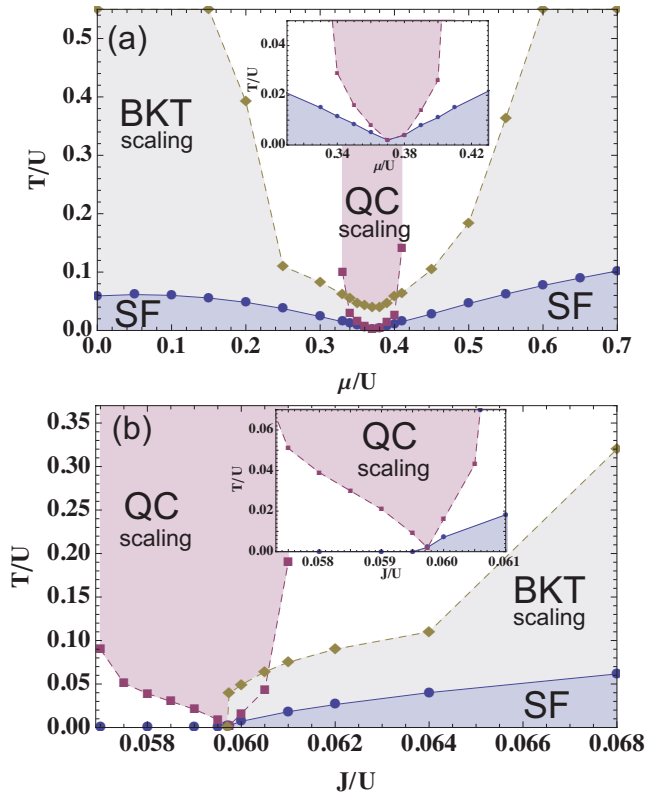}
\caption{
(a) Temperature scales as a function of chemical potential, $\mu/U$, at $J=J_c=0.05974 U$. Blue filled circles are the true phase transition boundary, $T_c$, while the brown diamonds indicate the upper bound for the BKT scaling regime, i.e. $T_\xi^{BKT}$. $\log (\xi)\propto (T-T_{c})^{-1/2}$ up to 99\% accuracy in the yellow shaded regime between above two curves. Red squares indicate $T_\xi^{QC}$, above which (pink regime) is the universal scaling regime ($\xi\propto 1/T$) near the quantum critical point.
(b) Temperature scales as a function of tunnelling amplitude, $J/U$, at $\mu=\mu_c=0.37$. The left hand side and right hand side of $J_c=0.05974 U$ are respectively Mott insulator and superfluid phases at zero temperature. All other notations are the same as in (a).
The regime near the quantum critical point is maximized in the insets, where we only show the $T_c$ and $T_\xi^{QC}$ for clearness. Note that the $T_c$ is shown slightly above $0$ at $(\mu, J)= (\mu_c, J_c)$ due to the numerical limitation.
The overlap of BKT regime and quantum critical regime near $J=J_c$ is simply due to the 1\% tolerance of the numerical fitting results, without any physical implication.
}

\label{fig:crossover2}
\end{figure}

In Fig. \ref{fig:crossover2}(a) and its inset, we show the calculated two crossover temperatures, $T_\xi^{BKT}$ and $T_\xi^{QC}$, as a function of chemical potential at $J=J_c$. It is surprising to find that the regime for the BKT scaling regime is much larger than the phase transition temperature (i.e. $T_\xi^{BKT}\gg T_c$) when slightly away from the quantum critical point. Besides, the quantum critical regime is symmetric along the horizontal axis (chemical potential), but also very narrow in the parameter regime (around $|u-\mu_c|/U\sim 0.03$). 
   
%---------------
%---------------

In Fig. \ref{fig:crossover2}(b), we further show the universal scaling regime near the superfluid-to-Mott insulator transition point. In the superfluid side, we again observe that $T^{BKT}_{\xi}$ is much larger than $T_c$ and increases when away from the tip of the quantum critical point, showing the overwhelming dominance of the classical BKT mechanism. On the other hand, the universal scaling regime near the quantum critical point, indicated by $T^{QC}_{\xi}$, increases sharply and exceeds the $T^{BKT}_\xi$ as departing from the QCP. In the disordered side, where the ground state is Mott insulator, $T^{QC}_{\xi}$ increases as a function $|J-J_c|/U$ in a much slower rate. 
%From the inset one can find that $T_\xi^{OC}\sim\xi^{-2}\sim |J-J_c|^{2\nu}$ in the disordered side. 
Such a strong asymmetric behavior of the quantum critical regime, results from the fact that the excitation spectrum in the superfluid side and the disordered (normal phase) are completely different. To the best of our knowledge, this is the first results obtained by a numerical calculation without any approximations. 

%According to the comparison between the $T^{QC}_{\xi}$ and the $T^{BKT}_{\xi}$, the range of quantum %critical regime is determined to be 
%$\mu/U\approx 0.37\pm 0.04$ and asymmetric $J/U\approx 0.05974^{+0.001}_{-0.002}$ from the quantum critical %point (at $\mu=\mu_c$ and $J=J_c$). %, as shown in the red shadow in the inset of Fig.\ref{fig:crossover2}%%(b). 
%Inside this QC regime, the temperature scale of $T^{QC}_{\xi}$ can also be up to $0.37U$ or more (depending %on the fitting criteria. 
%Furthermore, we find that $T^{BKT}_{\xi}$ can be much exceed the $T_c$ by about $0.8U$ in the deep %superfluid regime. 
%As a result, we can expect that the universal scaling behaviour near the critical point (either for BKT or %for QC regime) should be experimentally measurable in the present condition. 

%%%%%%%%%%%%%%%%%%%%%%%%%%%%%%%%
\section{The experiment-related issues}

Now we briefly discuss the experimentally related issue for measuring the critical regime.
For a normal fluid above the critical temperature, the correlation length, $\xi$, can be directly measured via matter wave interferometry \cite{Hadzibabic, Esslinger_Science, BlochNature2000, NavonScience} and/or {\it in situ} momentum distribution imaging \cite{JochimPRL2015}. It has been shown that the result is not sensitive to finite size effects.

\begin{figure}
\centering
\includegraphics[width=7.5cm]{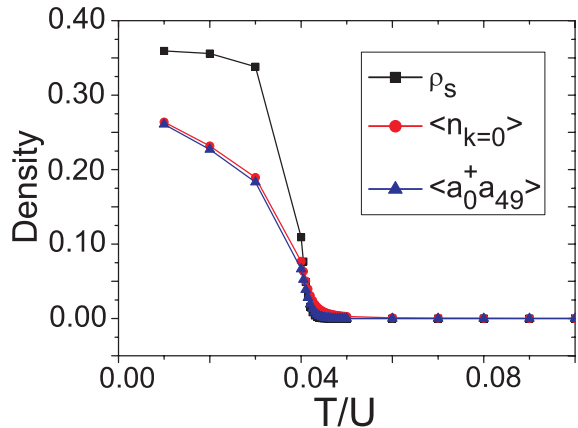}

\caption{
The superfluid density, condensate density and the $\left\langle a^\dagger_0 a_{L/2}\right\rangle$ as function of temperature at $\mu/U=0$ and $J/U=0.04$ in $100\times100$ square lattices.
%This figure quantitatively shows the differences of quasi-condensate and superfluid in $100\times100$ square lattice. 
The black squares are superfluid density obtained by measuring winding number, the red circles are the condensate density by taking $k=0$ in the Fourier transform of the correlation function, and blue triangles are $\left\langle a^{\dagger}_{0}a_{49}\right\rangle$.
%, which gives the evidence that the quasi-condensate density is mainly contributed by finite-size effect. 
%The horizontal axis is temperature, and the other parameters are same as Fig. \ref{fig:aa-r} which is far from quantum critical regime. 
%(b) Near quantum critical regime, the difference between quasi-condensate and superfluid becomes large. 
%The dash lines are quasi-condensate density, and the solid lines are superfluid density. 
%Various colors are different $J/U$ with $\mu/U=0.35$. $J/U=0.061$ for purple dots, and $0.06$ for blue dots.
}
\label{fig:QCSF-T}
\end{figure}

When the temperature is below the critical temperature (i.e. superfluid regime), the scaling exponent, $\eta$, can be measured from the condensate density through the finite size effect. It is because the condensate density, $\left\langle n_{k=0}\right\rangle=\sum_{i,j} G_{i,j}\propto\sum_j G_{0,j}$, is mostly dominated by the long distance behaviour since it decays slowly as a power law, i.e. $\left\langle n_{k=0} \right\rangle\cong G_{0,L/2}$. Thus, the experimental measurements of $\eta$ can be done by measuring $\langle n_{k=0}\rangle\propto (L/2)^{-\eta}$ through the time-of-flight experiments with different system sizes. In Fig. \ref{fig:QCSF-T}, we show the superfluid density measured by winding number (black squares), the condensate density by taking $k=0$ in the Fourier transform of the correlation function (red circles), and the correlation function of $G_{0,L/2}$ as function of temperature (blue triangles).
It is clear to see that both the condensate densities and $G_{0,L/2}$ shrink sharply to zero as $T$ increasing to $T_c$, and they are clearly very close to each other. 

Considering more realistic cases, like the systems of $^{133}$Cs or $^{87}$Rb in an optical lattice \cite{Chin_Nat, Ian_PRL105}, one can fix the tunnelling rate, $J$, and expect to see different regime in space by using local density approximation, i.e. the local chemical potential changes due to the inhomogeneous harmonic trap \cite{Fang_QMC}. From the results of Fig. \ref{fig:crossover2}(a), we estimate that the quantum critical regime (the pink area) may be observed in spatial dimension of approximately 7 sites $\times$ 7 sites in a typical trapping potential. On the other hand, recent development of the box trap systems in recent experiments also give an optimistic prospect in enlarging the system size beyond local density approximation \cite{HadzibabicBoxTrap,DalibardBoxTrap}. 

%----------
%------------

%\section{conclusion}
%\underline{\textit{Conclusion}}

\section{Conclusion}
In summary, we quantitatively study the critical behaviors of the correlation function, $G_{i,j}$ below and above $T_c$ along with investigating the divergence of correlation length $\xi$ and the critical exponents $\eta$ and $\nu$. We quantify the ranges of classical BKT and quantum critical regimes, and show the possibility to reach them in current experiments. Our results pave the way to quantitatively study the critical behavior in other strongly correlated system.

We acknowledge the invaluable discussions with Pochung Chen, Chung-Hou Chung, Cheng Chin, Nick P. Proukakis and Chia-Min Chung. This work is supported by MOST and NCTS. 

%\end{acknowledgments}
%%%%%%%%%%%%%%%%%%%%%%%%%%%%%%
%\section{Ref}


\begin{thebibliography}{99}


%Realization of Ultracold Systems:%
\bibitem{ReviewBloch1}
I. Bloch, J. Dalibard, and W. Zwerger, Rev. Mod. Phys. {\bf 80}, 885 (2008), 

\bibitem{ReviewBloch2}
I. Bloch, J. Dalibard, and S. Nascimb\`{e}ne, Nat. Phys. {\bf 8}, 267 (2012).

\bibitem{ReviewChin}
C. Chin, R. Grimm, P. Julienne, and E. Tiesinga, Rev. Mod. Phys. {\bf 82}, 1225 (2010).

\bibitem{MetIns}%Metal-insulator transitions
M. Imada, A. Fujimori, and Y. Tokura, Rev. Mod. Phys. {\bf 70}, 1039 (1998).

\bibitem{BlochNature2000}
I. Bloch, T. W. H\"{a}nsch and T. Esslinger, Nature {\bf 403}, 166 (2000).

\bibitem{JochimPRL2015}%paircondensate
M. G. Ries, A. N. Wenz, G. Z\"{u}rn, L. Bayha, I. Boettcher, D. Kedar, P. A. Murthy, M. Neidig, T. Lompe, and S. Jochim, Phys. Rev. Lett. {\bf 114}, 230401 (2015),
P. A. Murthy, I. Boettcher, L. Bayha, M. Holzmann, D. Kedar, M. Neidig, M. G. Ries, A. N. Wenz, G. Z\"{u}rn, and S. Jochim, {\it ibid.} {\bf 115}, 010401 (2015).

\bibitem{CMC}%paircondensate
C.-M. Chung, S. Fang, and P. Chen, Phys. Rev. B {\bf 85}, 214513 (2012).

\bibitem{Laloe}
M. Holzmann, G. Baym, J.-P. Blaizot, and F. Lalo\"{e}, Proc. Natl. Acad. Sci. {\bf 104}, 1476 (2007).

\bibitem{Smith}
R. P. Smith, N. Tammuz, R. L. D. Campbell, M. Holzmann, and Z. Hadzibabic, Phys. Rev. Lett. {\bf 107}, 190403 (2011).

\bibitem{Ohashi}
M. Matsumoto, D. Inotani, and Y. Ohashi, Phys. Rev. A {\bf 93}, 013619 (2016).

%\bibitem{Holthaus}
%D. Hinrichs, A. Pelster, and M. Holthaus, App. Phys. B {\bf 113} 57 (2013).

\bibitem{Mueller}
K. R. A. Hazzard and E. J. Mueller, Phys. Rev. A {\bf 84}, 013604 (2011).

\bibitem{ZhouHo}
Q. Zhou and T.-L. Ho, Phys. Rev. Lett. {\bf 105}, 245702 (2010).

%\bibitem{VicariPRL09} maybe too old
%Massimo Campostrini and Ettore Vicari, PRL 102, 240601 (2009).

\bibitem{VicariPRB14}
M. Campostrini, A. Pelissetto, and E. Vicari, Phys. Rev. B {\bf 89}, 094516 (2014).

%\bibitem{Krauth} maybe too old
%Markus Holzmann,1,* Maguelonne Chevallier,2 and Werner Krauth2,†, PHYSICAL REVIEW A 81, 043622 (2010).

\bibitem{Holzmann}
I. Boettcher and M. Holzmann, Phys. Rev. A {\bf 94}, 011602 (2016).

\bibitem{NavonScience}
N. Navon, A. L. Gaunt, R. P. Smith, and Z. Hadzibabic, Science {\bf 347}, 167 (2015).

\bibitem{ProkofevPRL14}%, critical conductivity
K. Chen, L. Liu, Y. Deng, L. Pollet, and N. Prokof’ev, Phys. Rev. Lett. {\bf 112}, 030402 (2014).

\bibitem{ChinNature11}
C.-L. Hung, X. Zhang, N. Gemelke1, and C. Chin, Nature {\bf 470}, 236 (2011).

\bibitem{Esslinger_Science}
T. Donner, S. Ritter, T. Bourdel, A. Öttl, M. K\"{o}hl, and T. Esslinger, Science {\bf 315}, 1556 (2007).

\bibitem{ChinCriticalityNJP}
X. Zhang, C.-L. Hung, S.-K. Tung, N. Gemelke, and C. Chin, New J. of Phys. {\bf 13} 045011 (2011).

\bibitem{BKT}
V. L. Berezinskii, Sov. Phys. JETP, {\bf 32}, 493 (1971), V. L. Berezinskii, {\it ibid}, {\bf 34}, 610 (1972), J. M. Kosterlitz and D. J. Thouless, J. Phys. C: Solid State Phys., {\bf 6}, 1181 (1973).

\bibitem{PathIntegral1987}
E. L. Pollock and D. M. Ceperley, Phys. Rev. B {\bf 36}, 8343 (1987).

\bibitem{Gora_QC}
Yu. Kagan, B. V. Svistunov, and G. V. Shlyapnikov, Sov. Phys. JETP {\bf 66}, 314 (1987).

\bibitem{Cooper}
N. R. Cooper and Z. Hadzibabic, Phys. Rev. Lett. {\bf 104}, 030401 (2010).

\bibitem{JSY_aniso} %low-D
J.-S. You, H. Lee, S. Fang, M. A. Cazalilla, and D.-W. Wang, Phys. Rev. A {\bf 86}, 043612 (2012).

\bibitem{Hadzibabic}
Z. Hadzibabic, P. Kr\"{u}ger, M. Cheneau, B. Battelier, and J. Dalibard, Nature {\bf 441}, 1118 (2006).

\bibitem{Phillip_QCvsSFexp}
P. Clad\'{e}, C. Ryu, A. Ramanathan, K. Helmerson, and W. D. Phillips, Phys. Rev. Lett {\bf 102}, 170401 (2009).

\bibitem{Ian_PRL105}
K. Jim\'{e}nez-Garc\'{i}a, R. L. Compton, Y.-J. Lin, W. D. Phillips, J. V. Porto, and I. B. Spielman, Phys. Rev. Lett. {\bf 105}, 110401 (2010).

\bibitem{Spielman_CondensateFraction}
I. B. Spielman, W. D. Phillips, and J. V. Porto, Phys. Rev. Lett. {\bf 98}, 080404 (2007) and {\bf 100}, 120402 (2008).

\bibitem{Chin_Nat}
N. Gemelke,  X. Zhang,  C.-L. Hung and C. Chin, Nature {\bf 460}, 995 (2009).

\bibitem{FisherFisher}
M. P. A. Fisher, P. B.Weichman, G. Grinstein, and D. S. Fisher, Phys. Rev. B {\bf 40}, 546 (1989).

\bibitem{Polkovnikov_PNAS}
A. Polkovnikov, E. Altman, and E. Demler, Proc. Natl. Acad. Sci. {\bf 103}, 6125 (2006).

\bibitem{Endres}
M. Endres, M. Cheneau, T. Fukuhara, C. Weitenberg, P. Schau\ss, C. Gross, L. Mazza, M. C. Banuls, L. Pollet, I. Bloch, S. Kuhr, App. Phys. B {\bf 113}, 27 (2013).

\bibitem{Lode_QMC}
L. Pollet, N. V. Prokof’ev, and B. V. Svistunov, Phys. Rev. Lett. {\bf 104}, 245705 (2010).

\bibitem{Fang_QMC}
S. Fang, C.-M. Chung, P.-N. Ma, P. Chen and D.-W. Wang, Phys. Rev. A {\bf 83}, 031605 (2011).

\bibitem{Rancon}
A. Ran\c{c}on and N. Dupuis, Europhys. Lett. {\bf 104}, 16002 (2013).

\bibitem{Prokofev1}
N. V. Prokof’ev, B. V. Svistunov, and I. S. Tupitsyn, Phys. Lett. A {\bf 238}, 253 (1998); Sov. Phys. JETP {\bf 87}, 310 (1998) [Zh. Eksp. Teor. Fiz. {\bf 114}, 570 (1998)].

\bibitem{Prokofev2}
N. V. Prokof’ev and B. V. Svistunov, Phys. Rev. Lett. {\bf 87}, 160601 (2001).

%\bibitem{Greiner}
%M. Greiner, O. Mandel, T. Esslinger, T. W. H\"{a}nsch, and I. Bloch, Nature {\bf 415}, 39 (2002).
%\bibitem{Trotzky}
%S. Trotzky, L. Pollet, F. Gerbier, U. Schnorrberger, I. Bloch, N. V. Prokof’ev, B. Svistunov, and M. Troyer, Nat. Phys. {\bf 6}, 998 (2010).
%\bibitem{Gerbier}
%F. Gerbier, S. Trotzky, S. F\"{o}lling, U. Schnorrberger, J. D. Thompson, A. Widera, I. Bloch, L. Pollet, M. Troyer, B. Capogrosso-Sansone, N. V. Prokof’ev, and B. V. Svistunov, Phys. Rev. Lett. {\bf 101}, 155303 (2008).

%\bibitem{Fang_tof}
%S. Fang, R.-K. Lee, and D.-W. Wang, Phys. Rev. A {\bf 82}, 031601 (2010).

%\bibitem{Chin_insitu}
%C.-L. Hung, X. Zhang, N. Gemelke, and C. Chin, Phys. Rev. Lett. {\bf 104}, 160403 (2010).
%Critical Behavior of a Trapped Interacting Bose Gas

%New Schemes:%
%\bibitem{Daley}
%A. J. Daley, H. Pichler, J. Schachenmayer, and P. Zoller, Phys. Rev. Lett. {\bf 109}, 020505 (2012).

\bibitem{MerminWagner}
N. D. Mermin and H. Wagner, Phys. Rev. Lett. {\bf 17}, 1133 (1966).

\bibitem{Zinn-Justin}  %Critical Behaviours Ref:%
J. Zinn-Justin, \textit{Quantum Field Theory and Critical Phenomena}, Fourth Edition, Oxford (2002).

%\bibitem{Chin}
%density-density correlation
%C.-L. Hung, X. Zhang, L.-C. Ha, S.-K. Tung, N. Gemelke, and C. Chin, New J. Phys. {\bf 13} 075019 (2011).






\bibitem{Sachdev}
S. Sachdev, \textit{Quantum Phase Transition}, Second Edition, Cambridge (2011).

\bibitem{Barbara_nu}
B. Capogrosso-Sansone, \c{S}. G. S\"{o}yler, N. Prokof’ev, and B. Svistunov, Phys. Rev. A {\bf 77}, 015602 (2008).

%\bibitem{Igor}
%I. Herbut, \textit{A Modern Approach to Critical Phenomena}, Cambridge (2010).
\bibitem{SachdevPRB}
S. Sachdev, Phys. Rev. B,  {\bf 55}, 142 (1997).
\bibitem{HadzibabicBoxTrap}
A. L. Gaunt, T. F. Schmidutz, I. Gotlibovych, R. P. Smith, and Z. Hadzibabic, Phys. Rev. Lett. {\bf 110}, 200406 (2013).
\bibitem{DalibardBoxTrap}
L. Chomaz, L. Corman, T. Bienaim\'{e}, R. Desbuquois, C. Weitenberg, S. Nascimb\`{e}ne, J. Beugnon, and J. Dalibard, Nature Comm.{\bf 6}, 6162 (2015).
%\bibitem{Lozovik}
%N. K. Kultanov and Yu. E. Lozovik, Physics Letters A,223,189(1996).
%\bibitem{Cserti}
% \href{}{}J. Cserti, Am.J.Phys.68:896-906,2000


\end{thebibliography}
\end{document}